# Zeolite-templated 3D printed carbon nanotube networks with enhanced mechanical properties

Rushikesh S. Ambekar[1][†], Eliezer F. Oliveira[2,3,4][†], Brijesh Kushwaha[1], Leonardo D. Machado[5], Mohammad Sajadi[2], Ray H. Baughman[6], Pulickel M. Ajayan[2], Ajit K. Roy[7], Douglas S. Galvao[3,4][*], Chandra S. Tiwary[1,2][*]

[1]Metallurgical and Materials Engineering, Indian Institute of Technology Kharagpur, Kharagpur-382355, India
[2]Department of Material Science and NanoEngineering, Rice University, Houston, Texas, 77005 United States
[3]Applied Physics Department, State University of Campinas (UNICAMP), Campinas, SP, Brazil
[4]Center for Computational Engineering & Sciences (CCES), State University of Campinas - UNICAMP, Campinas, SP, Brazil
[5]Department of Theoretical and Experimental Physics, Federal University of Rio Grande do Norte (UFRN), Natal, RN, Brazil
[6]Alan G. MacDiarmid NanoTech Institute, The University of Texas at Dallas, Dallas, Texas, 75080-3021, United States
[7]Materials and Manufacturing Directorate, Air Force Research Laboratory, Wright Patterson AFB, OH 45433-7718, United States
[†] Equal contribution

Email: chandra.tiwary@metal.iitkgp.ac.in (C.S.T) and galvao@ifi.unicamp.br (DSG)

**Abstract**

Specific strength (strength/density) is a crucial factor while designing high load-bearing architecture in areas of aerospace and defense. The strength of the material can be enhanced by blending it with high strength components and/or fillers, but both options have limitations such as at certain loads, the materials can still fail due to poor filler and matrix interactions. Therefore, there is a great interest in enhancing the strength of materials by playing with topology/geometry. In this work, we have investigated the mechanical properties of zeolite-templated carbon nanotube networks (CNTnets). The atomic models were used to generate macro models that were 3D printed. The mechanical properties of CNTnets were investigated through fully atomistic molecular dynamics simulations and load-bearing tests. Our results show that some aspects of mechanical behavior proved to be scale-independent. The 3D printed structures were able to support high compressive loads without structural failure. Such complex architectures can be exploited for ultralight aerospace and automotive parts.

**1. Introduction**

Structures for load-bearing applications have been made of different materials, such as metals, ceramics, and polymers. For a large variety of applications, polymers are still the most preferred material due to its low density, high specific strength, extrudability, and cost. On the other hand, metals have higher density and ceramics are brittle and have poor processability[1–3]. The load-bearing capability of the pristine polymers can be enhanced by blends and/or with interpenetrating networks, but the compatibility among polymers can limit the candidate selection[4]. Composites consist of a matrix (polymer resin) with reinforcements (mostly ceramic or metal), where the matrix helps to facilitate the processing and the reinforcements bear the loads, but at critical load slippage of reinforcement from the matrix can occur due to poor matrix-reinforcement interaction as matrix and reinforcement are different class of materials[5]. Due to these limitations, more recently the focus has shifted towards the design-oriented approach instead of the morphological strengthening approach[6].

In this more topology-oriented approach, the pore geometry and size are essential aspects of the design, as they are key factors which will determine the mechanical strength of the structure. Pore shapes such as cubic, spherical, cylindrical, octahedral, tetrahedral contribute to the structural advantage under load[7]. Some nature occurring materials exhibit high porosity in complex topologies, which have inspired biomimetic design like airplanes inspired by birds[8], design of Gherkin tower situated in London inspired by the venus flower basket[9], self-cleaning surfaces inspired by gecko feet[10], hydrophobic surfaces inspired by silver ragwort leaf and lotus leaf[11], artificial bone inspired by natural bone[12] and synthetic shark skin inspired by shark texture to reduce drag[13], among others. In spite of numerous successful biomimicry cases, its broad approach is difficult due to extreme structural complexity and the limitation imposed by material properties. A simpler and desirable approach would be to combine easy processable but retaining structural complexity, especially for enhanced load-bearing applications[14].

There are many techniques available to fabricate porous structures such as gas foaming[15], solvent casting/particulate leaching[16] and 3D printing, out of which 3D printing is one of the most promising for porous structure due to its ability to fabricate intricate structure, cost-effectiveness, scalability, controlled porosity (size and shape)[17]. Examples of successful cases of 3D printed materials along these lines are: Kaur *et al.*[18] fabricated cellular truss inspired structure via 3D printing technique. Three different materials such as nylon, poly(lactic acid) (PLA), and carbon fiber reinforced PLA (CFRPLA) were fabricated with an octet and octahedral pore geometry. The compression tests showed that CFRPLA octet structure (0.37 GPa) has higher Young's modulus than PLA octet structure (0.29 GPa), similarly, CFRPLA octahedral structure (0.58 GPa) has higher Young's modulus than PLA octahedral structure (0.45 GPa)[18]. Belhabib *et al.*[19]. developed a mathematical model inspired in acrylonitrile butadiene styrene (ABS) porous cellular structure. The ABS compression study reveals that varying pore density in the cellular structure, compression strength varies from 2.7 up to 6.3 MPa[19]. Naghieh *et al.*[20] reported poly (lactic acid)/gelatin-forsterite scaffolds via a combination of 3D printing and electrospinning technique. Effect of extruder temperature on the mechanical properties have been also studied, for instance, pristine PLA structure printed at 210˚C (0.12 GPa) has a higher elastic modulus than printed at 192˚C (0.11 GPa) temperature[20].

Recent studies have also shown that combining computational modeling with 3D printing technology is an effective way to generate new structures using geometric design principles[21,22]. This approach opens new and quite interesting perspectives to design new materials with enhanced mechanical properties. One recent example of this is the structural and mechanical properties of 3D printed macroscale (cm size) schwarzite crystals generated from fully atomistic models[21]. Schwarzites are 3D porous carbon-based structures idealized by Mackay and Terrones[23] that exhibit positive and negative curvature topologies with tunable porous size and shape, and interesting properties[24–29]. This study showed that the qualitative behavior observed in the mechanical properties of schwarzites structures from molecular dynamics (MD) simulations of atomistic models, such as deformation patterns, high failure strain, and Young's modulus values, agree very well with the observed ones from the experimentally 3D printed schwarzite macro scale models (printed using PLA). In this way, to generate corresponding macro-scale structures of atomic-scale models of structures that are almost impossible to synthesize with the present technologies has proven to be an effective approach to create new structures with enhanced structural and mechanical properties.

Carbon-based nanostructures with complex geometries have been intensively studied. Depending on the architectural geometry of these structures, remarkable mechanical and electrical properties are obtained[30–34]. As the studies of schwarzites have shown that carbon-based nanostructures with complex geometries can be used as models to generate 3D printed macroscale structures and that their properties follow similar patterns observed at the nanoscale[21], it will be interesting to look for other nanostructures to be used in the same way.

Recent studies have indicated that a feasible way to synthesize new carbon-based porous nanostructures is to use zeolites as templates[35–37]. Zeolites are, in general, aluminosilicates whose structures can contain several pores and channels of molecular sizes, that can be used as template to create porous nanostructures inside them. Thus, it will be interesting to study the properties of these possible formed structures inside the pores of the zeolites and select those that present desired mechanical properties to be 3D printed at macro-scale.

In this work, 3D printing technique, in particular, fused deposition modeling, was used to fabricate zeolite-template carbon nanotube nets (CNTnets) structures for load-bearing function. The atomic models were used to generate macro models that were 3D printed using PLA. The compression strength of PLA porous structure was investigated experimentally. The effect of varying pore density on compression strength was also investigated, as well as, the effect of layer deposition orientation on the compression behavior. These experimental data were contrasted against the ones obtained from MD simulations.

## 2. Materials and Method
### 2.1. Obtaining the beta zeolite-template models

Here we will give a short description of how the zeolite-based CNTnets were obtained; a more complete description can be found in reference[38]. Our structure models were template-based on the beta zeolite (BEA framework type, composed by Si and O atoms, $P4_122$ space group, and unit cell parameters a=b=12.63 Å, c=26.18 Å, α=β=ɣ=90° with 3840 atoms[39]. The BEA has two orthogonal and equivalent channels that partially intersect (along the y and z directions). To form CNTnets inside the BEA channels, first, we had to identify single-wall carbon nanotubes candidates that fit inside the BEA channels. We tested CNT with different chiral indices (n,m) inside the BEA channels and we carried out geometry optimization of the systems BEA+CNT in order to determine the configurations of the lowest energy. These calculations were carried out through molecular mechanics (MM) simulations using the well-known universal force field (UFF)[40], as implemented in the Materials Studio software[41]; an energy convergence tolerance of 0.001 kcal/mol and a force convergence tolerance of 0.5 kcal/mol/Å were used in order to assure well-converged structures. The size and shape of the unit cell were also optimized. In this step we used a large BEA periodic supercell with 12.3 nm along its channels, and the CNT lengths were fixed in 10 nm (their number of atoms depends on the CNT chirality). A large BEA supercell was used to prevent spurious inter-nanotube interactions. The CNT with chiralities (3,3), (4,2), (5,1), (6,0), (4,3), and (5,2) resulted in the most stables BEA+CNT structures (similar energies). The best results (in terms of minimum energy values) were obtained with the CNT (6,0), it was select to create the carbon networks.

To build the CNTnets, we need to create (energetically favorable) junctions between the chosen CNT. Two CNT (6,0) were inserted into the adjacent BEA channels (along the y and z directions). Non-bonded carbon atoms were then added into the space between the CNT and using MD simulations with the reactive ReaxFF force field[42] (as implemented in the computational package LAMMPS[43]), we let this system freely evolve until a junction between the CNT is spontaneously created. In this step, the positions of the CNT were kept fixed and various cycles of heating and cooling with a temperature range from 300 K to 1500 K were performed to produce a defect-less junction between the CNT. The temperature simulations are controlled using Nosé-Hoover thermostats, and the used timestep was of 0.2 fs. After these steps, an "X" type junction was obtained between the CNT, which is formed from joint heptagons and pentagons[44] (See Figure 1).

Deciding on which CNT and the junctions that would be employed to build the network structures, we have considered two different cases: one with partial filling of BEA channels (Structure 1 in Figure 1) and one with total filling of BEA channels (Structure 2 in Figure 1). These structures presented in Figure 1 were built from their replicated corresponding unit cells. The unit cell parameters of these two carbon network are 894 carbon atoms, mass density of 0.44 g/cm$_3$, a=28.47 Å, b=39.93 Å, c=39.22 Å, and α=β=ɤ=90° for Structure 1 and 321 carbon atoms, mass density of 1.28 g/cm$_3$, a=12.99 Å, b=13.68 Å, c=29.28 Å, and α=β=ɤ=90° for Structure 2. A thermal equilibration for temperatures of 300 K up to 600 K was also carried out for both structures in order to verify its thermal stability; both structures remained stable for the temperature range considered here.

### 2.2. Preparation of the structures for 3D printing

After obtaining the structures from MM/MD studies, we created structures to be printed and for the mechanical properties evaluation. Using VMD software[45], we create the structures 1 and 2 supercells of approximately the same volume of 20.10$_4$Å$_3$; these structures are presented in Figure 1. In order to compare the behavior of the BEA topology with the structures created from its channels filling, we replace its original atoms (Si and O) by carbon atoms. In this sense, we have both the positive and negative topological models of these structures. Using a similar procedure described before, we tested the stability of this carbon-based BEA and the structure 3 supercell (see Figure 1) was created with approximately the same size of the structures 1 and 2. The data of these carbon-based structures are presented in Table 1. Then, these structures were exported to a stereolithography file (STL) to be 3D printed.

Then, the obtained structures were fed in the FDM printer as a G-Code program; front view and 3D view of the molecular structure are shown in **Figure 1 (a)** and **Figure 1 (c),** respectively.

The designed structures were successfully fabricated in Flashforge adventure 3 printer, the front view and 3D view of 3D printed structures are illustrated in **Figure 1 (b)** and **Figure (d),** respectively. The precise dimensions of these fabricated structures are given in **Table 1**.

The commercial-grade of solid PLA filament was fed into the printer which is then subsequently heated in the extruder at 210˚C and extruded through a nozzle on the heated bed at 50˚C during printing. The paper tape was used for better adhesion of the structures to the printer bed, as fabricated structures have a single-layer thickness of 180 µm along the z-direction with 100% fill density. PLA filament (white) was provided by Flashforge 3DTechnology Co. Ltd. (Zhejiang, China) and has a diameter of 1.75 mm (with tolerance ±0.1), mass density of 1.210-1.430 gm/cm$_3$ and melting point 190˚C to 220˚C.

### 2.3. Compressive mechanical tests

Due to the structures be equal along the y and z directions (see Figure 1), the load-bearing capacity of the post-fabricated structures was tested along the x and z directions using the UTM SHIMADZU (AG 5000G), with a compression displacement rate maintained at 1 mm/min.

For qualitative comparison, we also performed the compressive tests in the molecular Structures 1, 2 and 3 using MD simulations. To do so, the finite structures were placed atop on a fixed 12-6 Lennard-Jones wall ($\varepsilon$ = 0.07 kcal/mol and $\sigma$ = 3.55 Å) and above them, another movable was placed. With the movable wall, a compressive uniaxial strain was applied to deform the structure and evaluate its mechanical response with a strain rate of -10$_{-6}$fs$_{-1}$. In order to decrease the thermal contribution in our deformation tests, these simulations were carried out at 10 K in an NVT ensemble. The stress-strain curves were obtained using the virial stress tensor component and the engineering strain at each compressive direction. In order to determine the spatially concentrated stress during simulations, we evaluated it through the von Mises stress, the second invariant of the deviatoric stress tensor[46]. This analysis also provides useful insights into the fracture dynamics. All deformation tests were performed using the ReaxFF force field[42], with a timestep of 0.1fs.

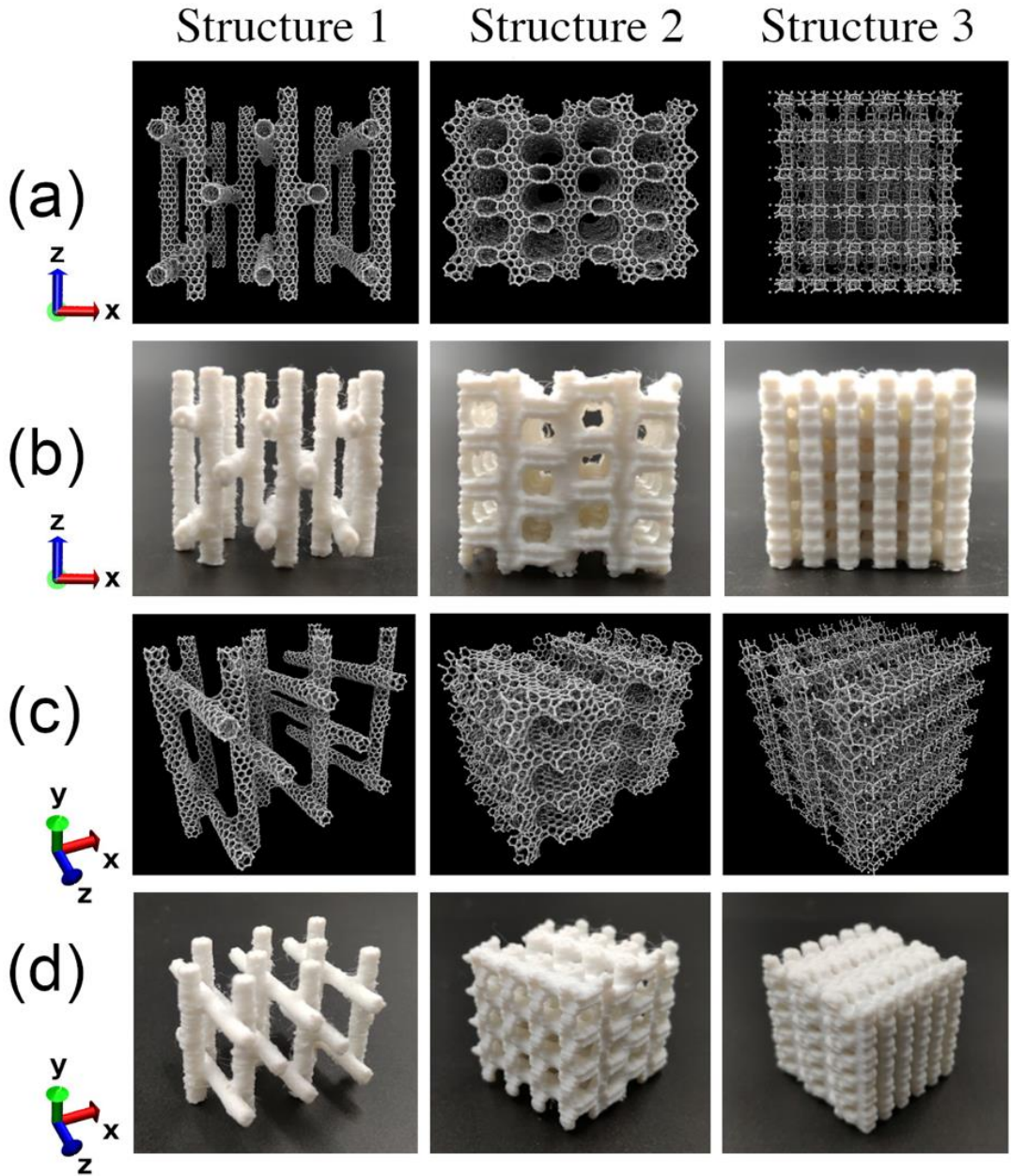

**Figure 1.** a) Front view of zeolite-templated molecular structures. b) Front view of 3D printed zeolite-inspired structures. c) 3D view of zeolite-inspired molecular structures. d) 3D view of 3D printed zeolite-inspired structures.

**Table 1.** Specifications of the created molecular structures from MD simulations and the 3D printed corresponding ones.

|  | **Theoretical MD** |  | **3D Printed** |  |
| --- | --- | --- | --- | --- |
| **Sample Name** | **Number of carbon atoms** | **Dimensions (Å)** | **Dimensions (cm)** | **Weight (gm)** |
| **Structure-1** | 16,944 | 78 x 75 x 71 | 3.0 x 2.955 x 3.044 | 17.560 |
| **Structure-2** | 10,799 | 62 x 50 x 56 | 3.0 x 2.994 x 2.869 | 11.5565 |
| **Structure-3** | 6,310 | 78 x 62 x 59 | 3.0 x 2.615 x 2.646 | 3.002 |

3. **Results and Discussion**

In order to discuss the mechanical behavior of the structures 1, 2 and 3, we will first present the MD results. In Figure 2 we present the stress-strain curves for compressive tests conducted along the x (Figure 2(a)) and z-direction (Figure 2(b)) for structures 1, 2, and 3. In Table 2 we present a summary of some mechanical properties. The structures 1, 2, and 3 have porosities of, respectively, 47.52%, 63.83%, and 88.42%, respectively. As can be seen from Figure 2, the curves representing the compressive tests along both x and z-directions present a similar behavior regarding the stress values, in which the stress values are inversely proportional to structural porosity. This indicates that the porosity of the structures, as expected, influences the stress level supported by them. In Figure 2(a)/x-direction, we observe a relatively small elastic region for the structures 2 and 3, in which a yield strength and stress of 3.7GPa and 1.2 % and 4.8 GPa and 0.8

% are observed, respectively. As for structure 1, the most porous one, a remarkable long elastic regime, with a yield strength of 0.6 GPa for a strain around 60%. These results suggest that the porosity play a crucial role regarding the deformation of the structure: the less porous structure, the shorter is its elastic regime and more stress is necessary to cause some deformation and achieve the plastic regime of the structure.

In Figure 2(b) we present the corresponding results for the z-direction. As can see from this Figure the shape of the stress-strain curves of compressive structures 2 and 3 are similar to the ones found for the x-direction, but the yield strength and strain values are quite different. We found a yield strength and strain for compressive tests for the z-direction of 3.6 GPa and 4.2 % and 3.9 GPa and 1.1 %, respectively, for structures 2 and 3. These values are lower than the corresponding ones for the x-direction. As for structure 1, a different behavior is seen in Figure 2(b) in comparison to Figure 2(a). Structure 1 does not present a significant elastic behavior along the z-direction, in which yield strength of 1.8 GPa at around 2.4 % of strain is found. These results indicate that beyond the porosity of the structure, the relative tube orientation significantly affects the observed mechanical properties. In general, we can see that the direction that is perpendicular to the nanotubes arrangements, which is the x-direction, presents the toughest and more elastic structures with decreasing porosity. Similar behavior was reported to other CNT nets[47], and the reasons are that is much more difficult to deform the tubes vertically than radially.

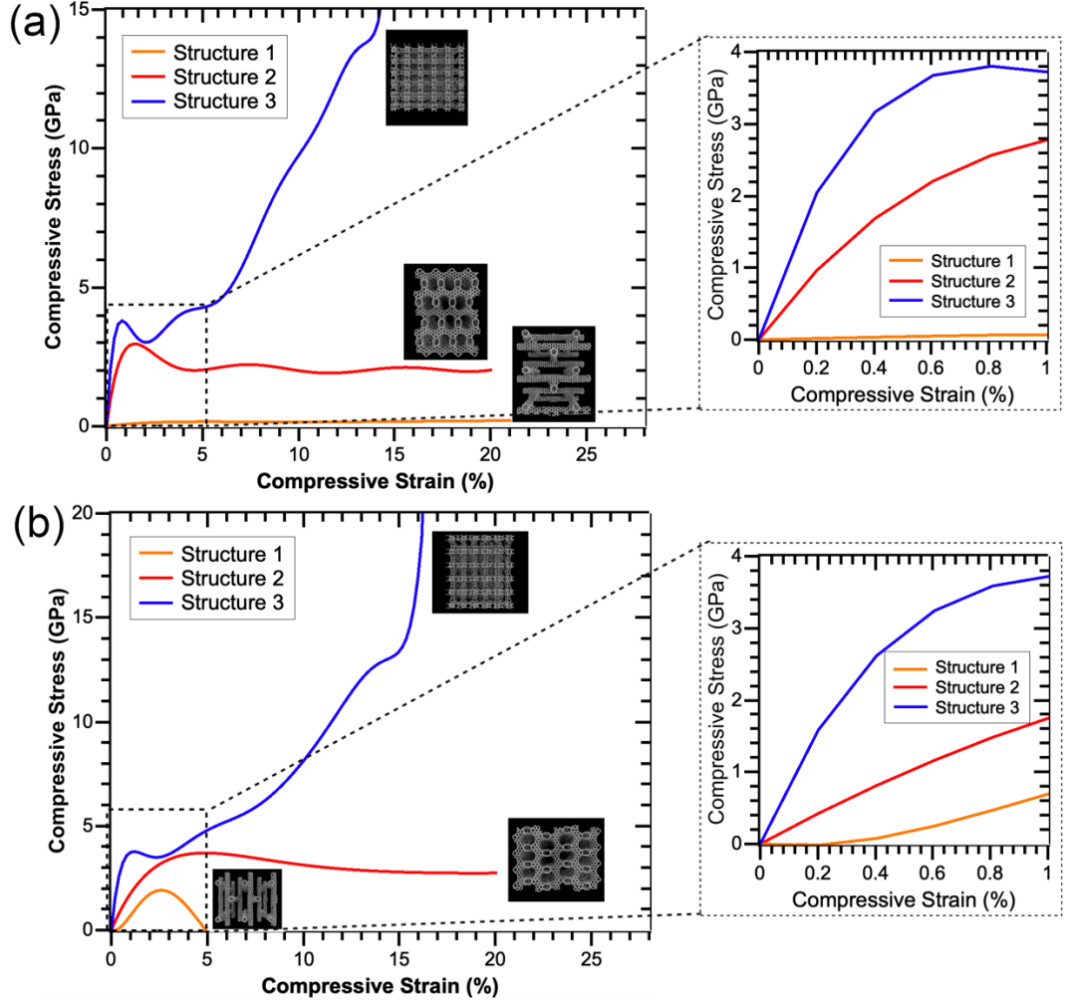

**Figure 2.** Stress-strain curves from MD simulations of structures 1, 2, and 3 for the compressive tests along (a) x-direction and (b) z-direction. The insets show the zoomed region of the initial compressive regimes.

The stress-strain curves of the PLA printed structures are presented in Figure 3. Comparing these experimental data with the corresponding ones from MD simulations presented in Figure 2, we observe a good qualitative agreement. This is an indication that some of the mechanical properties are scale-independent and mainly determined by the structural topology. This has been observed before for schwarzites[21] and tubulanes (3D network of cross-linked CNT)[47]. In this way, using atomic-scale structures to be 3D printed at macro-scale seems to be a good strategy to engineer new macro structures for load-bearing applications.

In Figure 3(a) we present the experimental stress-strain curves for the compressive tests of structures 1, 2, and 3 along the x-direction. Following the same trends from MD simulations, we observed that for the 3D print structures, the load capacity is inversely proportional to their porosity. Is should be remarked that in the present work, we have fabricated structures with a fused deposition modeling type 3D printer, therefore, polymeric chains are oriented along the x and y-direction, which are responsible for the anisotropic nature of as-prepared structures. The printed structures are susceptible to early failure when a compressive load is applied along the x-direction in comparison to the z-direction for the same reasons.

In Figure 4(b) we present the corresponding for z-direction, the compressive load is transversely applied on the oriented polymeric chains, therefore, the structure can accommodate higher load before deformations. The applied compression stress densifies the structure over the period of deformation, therefore, the strength of structure increases as the pores are deformed. In **Table 2** we present a summary of some of the mechanical properties.

The effect of the structural porosity on the Young's modulus and yield strength values are clear for both x and z-direction. The structure-1 has the highest porosity of 88.4% and therefore it has lower compression strength than structure 2 and 3. Young's modulus values of structure 1 along z-direction (89.3 MPa) is larger than structure-1 for the x-direction (9.8 MPa). For the x-direction, structure 3 (202.36 MPa) has a higher Young's modulus value than structure 2 (162.88 MPa); similarly, for the z-direction structure 3 (190.39 MPa) has a higher Young's modulus value than structure 2 (92.159 MPa), structure 2 (63.83%) has higher porosity than structure 3 (47.52%).

Yield strength of structure 2, structure 3 for the x-direction and structure 2 and structure 3 for the z-direction are 7.60, 11.20, 3.82, 9.52 MPa, respectively.

It was also observed that the relative structural tubular orientation has a significant effect on toughness, i.e., porous structures tested along the x-direction have 50% reduced toughness in comparison to the z-direction ones. The toughness of structure 2 (380KJ/m$_3$) and structure 3 (1045 KJ/m$_3$) for the z-direction are larger than the ones for structure 2 (200 KJ/m$_3$) and 3 (540 J/m$_3$) for the x-direction. As can be inferred from **Table 2**, there is again a good qualitative comparison between the experimental and theoretical data for the mechanical properties, thus validating the topology seems to be the fundamental feature defining the mechanical behavior.

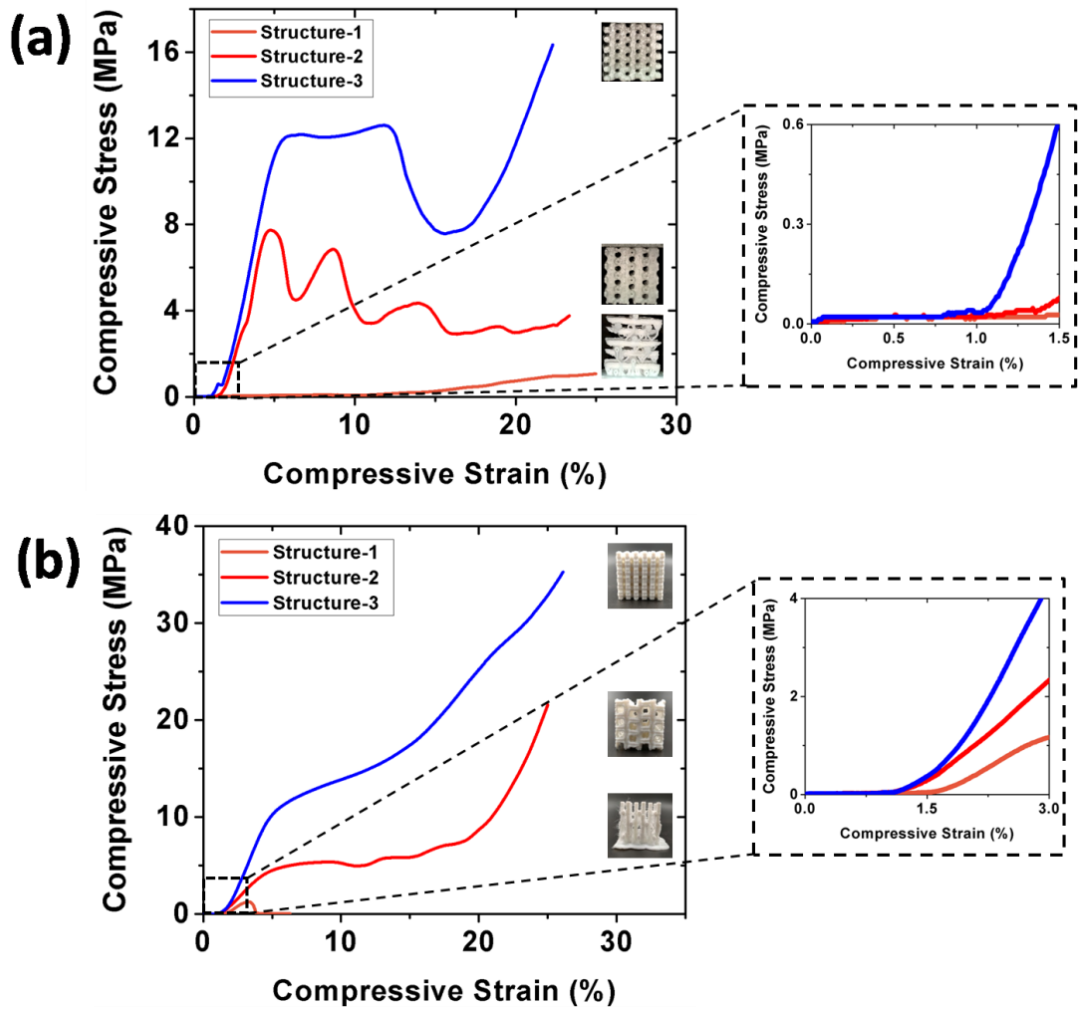

**Figure 3.** Experimental stress-strain curves for compressive tests of structures 1, 2, and 3 along (a) x-direction and (b) z-direction. The insets show the zoomed region of the initial compressive regimes.

**Table 2.** Mechanical properties data derived from experimental and theoretical compressive tests. The MD results are presented into parentheses.

| Orientation | x-direction | | | z-direction | | |
|---|---|---|---|---|---|---|
| **Structure** | 1 | 2 | 3 | 1 | 2 | 3 |
| **Porosity (%)** | 88.4 | 63.8 | 47.2 | 88.4 | 63.8 | 47.2 |
| **Young's Modulus** | 9.8 MPa (6.9 GPa) | 162.9 MPa (214.1 GPa) | 202.4 MPa (261.8 GPa) | 89.3 MPa (80.0 GPa) | 92.2 MPa (160.1 GPa) | 190.4 MPa (277.6 GPa) |
| **Yield Strength** | 0.99 MPa (0.6 GPa) | 7.6 MPa (3.7 GPa) | 11.2 MPa (4.8 GPa) | 1.24 MPa (1.8 GPa) | 3.8 MPa (3.6 GPa) | 9.5 MPa (3.9 GPa) |
| **Yield Strain** | 23.78 % (60.0 %) | 4.5% (1.2%) | 5.0% (0.8%) | 3.27 % (2.4 %) | 4.1% (4.2 %) | 4.6% (1.1 %) |
| **Toughness (KJ/m$_3$)** | 73.1 | 200 | 540 | 12 | 380 | 1045 |

In order to better understand the load-bearing behavior of structures at atomic and macroscopic levels, we selected snapshots of the porous structures at some selected compression strain percentages during the compression tests. These snapshots are presented in **Figures** 4 and 5, for the x and z-directions, respectively. The theoretical MD snapshots are colored accordingly to the distribution of the von Mises stress values; the local stress level is indicated on the stress level bar.

The results for the x-direction clearly show a non-uniform compression of pores and the crack formation/propagation can occur between two printed layers, which cause poor load distribution (as shown in **Figure 4**). The MD results indicate that the stress is mostly accumulated at the nanotube junctions, and the failure of the structure occurs with the collapse of the pores and nanotubes. As can be seen, the high elasticity of the structure 1 for x-direction can be related to the easy deformability of the nanotubes. As this structure have a very high porosity and few tube junctions, the deformed tubes can be accommodated at the existing empty spaces without a structural collapse.

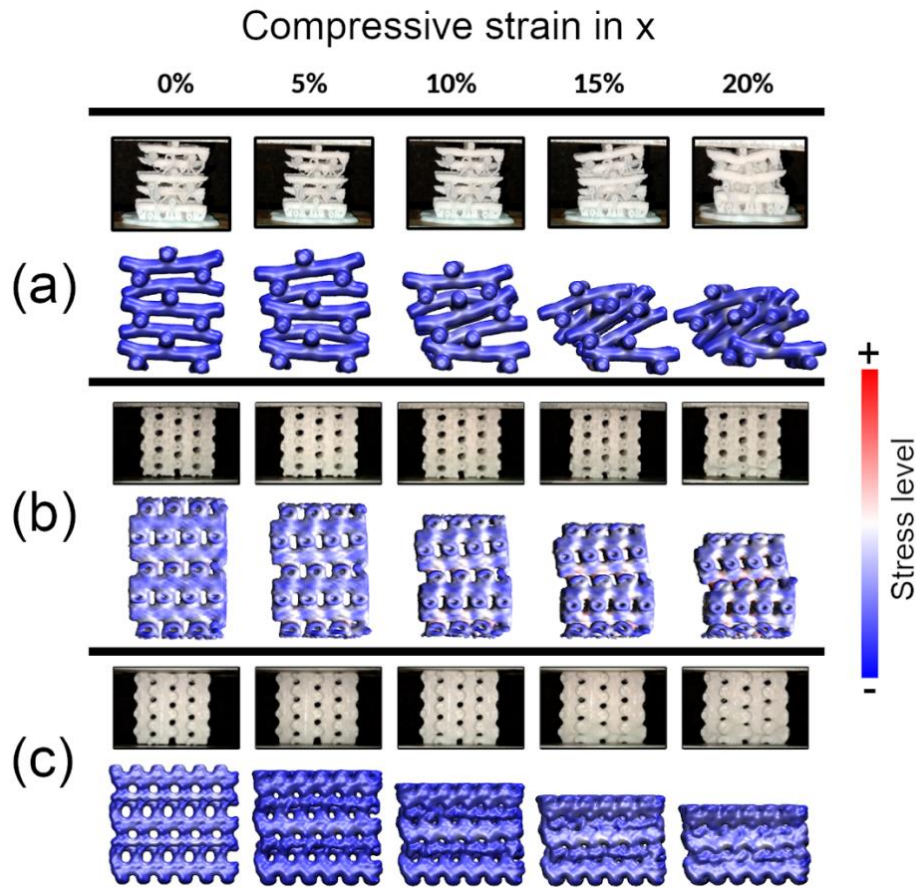

**Figure 4.** Snapshots of experimental and MD compressive tests for the x-direction of structures (a) 1,(b) 2, and(c) 3. For the MD ones, they are presented with the distribution of the von Mises stress values (accordingly to the color bar of stress level).

The structures tested along the z-direction exhibit a more uniform compression due to the compression load applied to the transverse structural orientation (as shown in **Figure 5**).

Differently from the observed for the x-direction, the most stressed part of the structures are now the nanotubes, that are arranged parallelly to the compressive direction. The failure of the structures in the z-direction are mostly related to the nanotube bending.

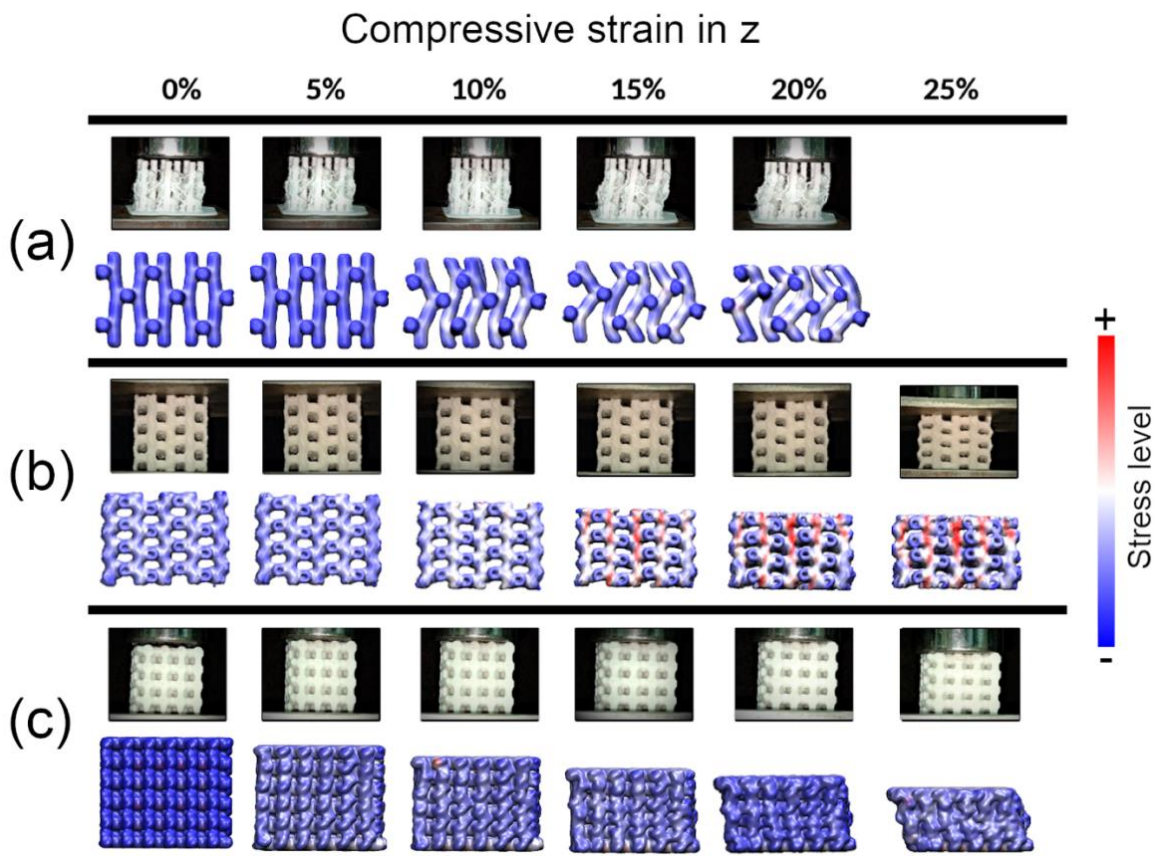

**Figure 5.** Snapshots of experimental and MD compressive tests for the z-direction of structures (a) 1,(b) 2, and(c) 3. For the MD ones, they are presented with the distribution of the von Mises stress values (accordingly to the color bar of stress level).

To have a better idea of how the deformations and failures occur, it is interesting to analyze the characteristics and behavior of the resulting templated structures from the BEA zeolite. As discussed before, the structures that can be created from BEA are basically a 3D network of joint carbon nanotubes. When these structures are printed into macroscale, the resulting structures can be regarded as a fused cylindrical structure, as shown in **Figure 6(a)**. When the templated structures are submitted to compressive stress, if they were not joined, the cylinders/nanotubes would be free to roll and/or to slide in order to accommodate the stress (see **Figure 6(a)**). As the cylinders/nanotubes are joined, the joins add resistance to roll and/or slide, which results in a stress accumulation in the joint regions.

The number of joints also determines the degree of freedom of the cylinders/nanotubes. Fewer number of joints implies that the more stress will be distributed along the structure due to the cylinders/nanotubes rolling and/or sliding. As can be seen in **Figures 6(b)** and **6(c)**, in which a representative part of the structures is shown, compression perpendicular to cylinders/nanotubes (x-direction) tends to make them to rotate and to bend in order to accommodate the stress. In the case of high-density structure (structure 2), the number of joints is higher, which decreases the rolling/slipping freedom of the cylinders/nanotubes; this results in smaller structural movements along the perpendicular direction to the compressive one. Then, the stress becomes more concentrated at the junctions and the stress yield is reached when the nanotubes can no longer roll and slip, which leads to the collapse of the cylinder/nanotube (see **Figure 6(c)**). As for the low-density structure (Structure 1), the number of joints is smaller, which makes the structure with more freedom to accommodate new conformations as the stress is increased. Because of this, when this structure is compressed along the x-direction, the rolling and slipping process allow it to spread as we compress it (see **Figure 6(a)**), structural nanotube collapse was not observed, mainly tube bending when the cylinders/nanotubes rolls (**Figure 6(b)**).

When the compression is applied parallel to cylinders/nanotubes (z-direction), it is difficult for the cylinders/nanotubes to roll and slip, which leads to a stress accumulation at parallel portions to the compressive direction, i.e., in the cylinders/nanotubes. Initially, the joints concentrate more stress than the cylinders/nanotubes, since they are trying to roll (start a pull-push process at the junctions in a tentative to roll). Then, as seen in **Figures 6(b)** and **6(c)**, the stress tends to dissipate perpendicular to the junctions. Because the junctions can withstand high-stress levels without breaking, the same does not happen to the nanotubes, they start to bend, which leads to the structure to collapse. Consequently, in compressions perpendicular to the cylinders/nanotube, the failure of the structures is mainly related to the bending yield.

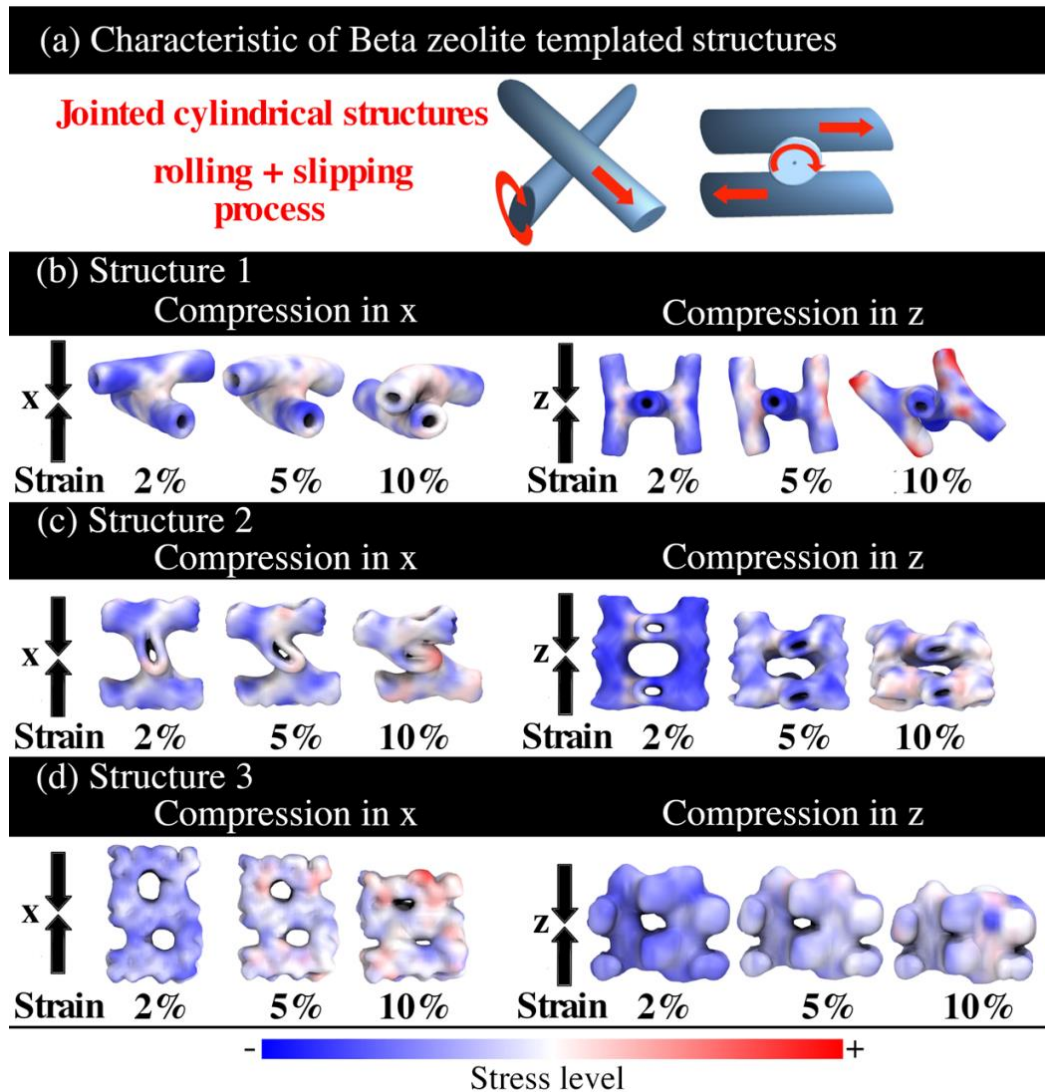

**Figure 6.** (a) Characteristics of beta zeolite template structure and compression snapshots at junctions of the structure-1, (c) structure-2 and (d) structure-3 along x and z-directions.

For the beta zeolite structure (structure 3 in Figure 1), we can see that it resembles an interpenetrated quasi-parallelepiped network with quasi-circular porous. Compared to structures 1 and 2, structure 3 has a significantly higher number of junctions and more mass. When structure 3 is compressed along the x- or z-directions, the stress is more delocalized along the structure (**Figures 4(c) and 5(c)**). In **Figure 6(d)**, it is possible to see that the failure of the structure occurs when the quasi-paralepidid structure fails around the zeolite channel, leading to a global structural collapse. This happens due to the quasi-paralepidid building block are rigid, making difficult the rolling and slipping movements.

## 4. Conclusions

In this work, we have investigated the mechanical properties of zeolite-templated carbon nanotube networks (CNTnets). The atomic models were used to generate macro models that were 3D printed. The mechanical properties of CNTnets were investigated through fully atomistic molecular dynamics simulations and load-bearing tests. There is a good qualitative agreement for the mechanical behavior of the atomic models and the corresponding 3D printed macro-scale ones. Some of the discrepancies are due to the intrinsic 3D printed layer-by-layer process. Our results also show that some aspects of mechanical behavior proved to be scale-independent. The 3D printed structures were able to support high compressive loads without structural failure. These conclusions are similar to other structures obtained with the same approach[21,47] and validate that generating 3D printed macro-scale structures of atomic-scale models of structures that are almost impossible to synthesize with the present technologies, is an effective approach to create new structures with enhanced structural and mechanical properties. Such complex architectures can be exploited for ultralight aerospace and automotive parts.

## 5. Acknowledgements

EFO and DSG thank the Brazilian agencies CNPq and FAPESP (Grants 2013/08293-7, 2016/18499-0, and 2019/07157-9) for financial support. Computational support from the Center for Computational Engineering and Sciences at Unicamp through the FAPESP/CEPID, the Center for Scientific Computing (NCC/GridUNESP) of São Paulo State University (UNESP), and the High-Performance Computing Center at UFRN (NPAD/UFRN) are also acknowledged. This study was financed in part by the Coordenação de Aperfeiçoamento de Pessoal de Nível Superior - Brasil (CAPES) –Finance Code 001.